\documentclass[10pt]{article} 
\usepackage{amssymb}
\usepackage{amsmath}
\usepackage{amsfonts}
\usepackage{amsthm}
\usepackage{latexsym} 
\usepackage[dvips]{epsfig}

\theoremstyle{plain}
 
\newtheorem{lemma}{Lemma}
\newtheorem{theorem}{Theorem}
\newtheorem*{assumptions}{Assumptions}
\newtheorem*{conjecture}{Conjecture}

\newtheorem*{main}{Theorem}

\newtheorem*{proposal}{Proposal}

\def\O{\mathcal{O}}

\font\SYM=msbm10 
\newcommand{\Real}{\mbox{\SYM R}}
\newcommand{\Complex}{\mbox{\SYM C}}

\font\tenscr=rsfs10 scaled1100
\font\sevenscr=rsfs7 
\font\fivescr=rsfs5 
\skewchar\tenscr='177 
\skewchar\sevenscr='177
\skewchar\fivescr='177 
\newfam\scrfam 
\textfont\scrfam=\tenscr
\scriptfont\scrfam=\sevenscr 
\scriptscriptfont\scrfam=\fivescr

\def\scri{{\fam\scrfam I}}

\begin{document}

\bibliographystyle{/home/gast/jav2/tex/reporthack}

\title{Does asymptotic simplicity allow for radiation near spatial infinity?}

\author{Juan Antonio Valiente Kroon \thanks{E-mail address:
 {\tt jav@ap.univie.ac.at}} \\
 Institut f\"ur Theoretische Physik,\\ Universit\"at Wien,\\
Boltzmanngasse 5, A-1090 Wien,\\ Austria.}

\maketitle

\begin{abstract}
A representation of spatial infinity based on the properties of
conformal geodesics is used to obtain asymptotic expansions of the
gravitational field near the region where null infinity touches
spatial infinity. These expansions show that generic time symmetric
initial data with an analytic conformal metric at 
infinity will give rise to developments with a certain type of
logarithmic singularities at the points where null infinity and
spatial infinity meet. These logarithmic singularities produce a
non-smooth null infinity. The sources of the logarithmic singularities
are traced back down to the initial data. It is shown that if the parts
of the initial data responsible for the non-regular behaviour of the
solutions are not present, then the initial data is static to a
certain order. On the basis of these results it is conjectured that
the only time symmetric data sets with developments having a smooth
null infinity are those which are static in a neighbourhood
of infinity. This conjecture generalises a previous conjecture
regarding time symmetric, conformally flat data. The relation of these
conjectures to Penrose's proposal for the description of the
asymptotic gravitational field of isolated bodies is discussed.
\end{abstract}


\section{Introduction}

A central issue in the general relativistic theory of isolated systems
is that of the verification of the so-called \emph{Penrose proposal}
\cite{Pen65a}. Following Friedrich \cite{Fri99,Fri03a}, the proposal
can be formulated as:
\begin{proposal}[Penrose, 1965]
Far fields of isolated systems behave like asymptotically simple
spacetimes in the sense that they can be smoothly extended to null
infinity after a suitable conformal rescaling.
\end{proposal}
That this proposal is not empty ---i.e. that there are examples of
radiative spacetimes--- is known thanks to work by Chru\'sciel \&
Delay \cite{ChrDel02}.  The idea behind their result was to combine a
modification of the gluing construction of Corvino \cite{Cor00}, which
yields initial data which are Schwarzschildean outside a compact set,
with the semiglobal hyperboloidal existence result of Friedrich
\cite{Fri86b}.
\begin{figure}[t]
\centering
\includegraphics[width=.4\textwidth]{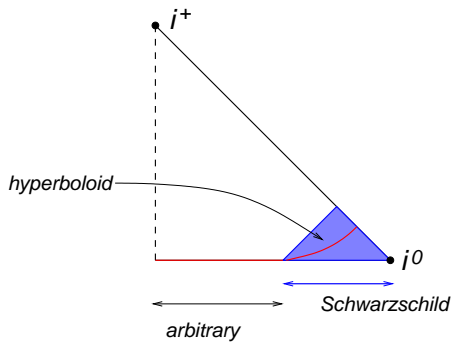} 
\caption{Schematic representation of the asymptotically simple
spacetimes constructed by Chru\'sciel \& Delay. The initial data is
Schwarzschildean outside a compact set.}
\label{chrusciel:fig}
\end{figure}
The resulting spacetime is then ---see figure \ref{chrusciel:fig}---
Schwarzschildean in a region, $\widetilde{N}$, of spacetime ``near
null and spatial infinity''. However, at later times radiation could
certainly be registered at null infinity. As emphasized by Friedrich,
the radiation content of this spacetime is rather special: the
Schwarzschildean nature of the initial data implies that
Newman-Penrose constants of the spacetime are zero; on the other hand,
the Newman-Penrose constants are know to be equal to the value of the
rescaled Weyl tensor at timelike infinity $i^+$
\cite{NewPen68,FriSch87}. Hence, the Weyl tensor at $i^+$ vanishes for
the Chru\'sciel-Delay spacetimes, thus implying a fast decaying 
gravitational field at very late times.

In the light of the results by Chru\'sciel \& Delay it is natural to
ask how general can be the behaviour of initial data in a
neighbourhood of spatial infinity if one is to obtain asymptotically
simple solutions. A first insight to this point was provided by the
analysis of Friedrich given in \cite{Fri98a}. In order to keep the
discussions in a reasonable complexity level, his analysis was
restricted to time symmetric initial data with an analytic
compactification at infinity. He found that a necessary condition in
order to attain spacetimes that are smoothly extendible at null
infinity is that the regularity condition
\begin{equation}
\label{reg_cond}
D_{(a_sb_s}\cdots D_{a_1b_1} b_{abcd)}(i)=0, \quad s=0,1,\ldots,
\end{equation}
 is satisfied to all orders. If the condition is not satisfied at some
 order, then the solutions to the Einstein field equations will
 develop logarithmic singularities at the ``sets where null infinity
 touches spatial infinity''. It is worth making a couple of remarks about
 this condition: firstly, that it is a purely asymptotic condition,
 and thus it does not pose a big limitation on the
 kind of physical systems one would like to describe. Secondly, it was
 proved in \cite{Fri88} that static initial data satisfies the
 regularity condition.

Recent work by the author \cite{Val04a}, using the techniques
developed by Friedrich in \cite{Fri98a} has shown that the regularity
condition (\ref{reg_cond}) is not a sufficient condition to ensure the
smoothness of the null infinity arising from developments of the class
of initial data under consideration ---time symmetric and with an analytic
initial 3-metric near infinity. The analysis in \cite{Val04a}
considered asymptotic expansions near null and spatial infinity
arising from time symmetric data which were further assumed to be
conformally flat near infinity. The assumption of conformal flatness
near infinity is in this context rather natural because it satisfies
automatically the regularity condition (\ref{reg_cond}). The results in
\cite{Val04a} lead way to the following conjecture:

\begin{conjecture}[Conformally flat data]
For every $k>0$ there exists a $p=p(k)$ such that the time development
of an asymptotically Euclidean, time symmetric, conformally smooth
initial data set which is conformally flat in a neighbourhood $B_a(i)$
of infinity admits a conformal extension to null infinity of class $C^k$
near spacelike infinity, if and only if the initial data set is
Schwarzschildean to order $p(k)$ in $B_a(i)$. Moreover, if the
conformal extension of the data is smooth ---i.e of class
$C^\infty$--- the data are exactly Schwarzschildean in $B_a(i)$.
\end{conjecture}

This conjecture, if found to be true, could be regarded as a rigidity
 result.  It is the objective of the present article to analyse what
 happens when the assumption of conformal flatness near infinity is
 removed from the initial data.  It could well be the case that the
 rigidity at spatial infinity suggested by the conjecture is an
 artifact of the ``specialness'' of the conformal
 flatness at infinity. It will be shown that this does not seem to be
 the case. Our main result ---cf. with the main theorem of
 \cite{Val04a}--- is the following.

\begin{main}[Main theorem]
A necessary condition for the development of time symmetric initial
data to be smooth at the intersection of null infinity and spatial
infinity is that the data set is at least static to order 3.
\end{main}

What we mean by being static up to an order $p$ will be clarified in
the main text. A more technical statement of the theorem will be also
be given there. Based on our main theorem, and noting that the only
time symmetric data which are conformally flat and static near
infinity are the Schwarzschildean ones, we put forward the following
conjecture:

\begin{conjecture}[General conjecture for time symmetric data]
For every $k>0$ there exists a $p=p(k)$ such that the time development
of an asymptotically Euclidean, time symmetric data set which is
conformally smooth in a neighbourhood $B_a(i)$ of infinity admits a
conformal extension to null infinity of class $C^k$ near spacelike
infinity, if and only if the initial data set is static to order
$p(k)$ in $B_a(i)$. Moreover, if the conformal extension is of class
$C^\infty$, then the data are exactly static in $B_a(i)$.
\end{conjecture}

As stated, the conjecture excludes totally the presence of
gravitational radiation near spatial infinity. Whether the latter is a
severe limitation for the modeling of systems of physical interest
remains to be seen. With regard to non-time symmetric data, one should
not expect the things to be any better. In this case it has been
shown that, for example, the presence of linear momentum in the
initial data produces solutions of the constraints which contain
logarithms ---see \cite{DaiFri01}. Analogous terms are, in the linear
case, source of further non-smoothness at the intersection of null and
spatial infinity ---see e.g. \cite{Val03a}. In any case, a good picture
of the complications of considering non-time symmetric data is not yet
available ---see however the results given in \cite{Val04b}.

This article is a natural continuation of the investigations
undertaken in \cite{Fri98a,FriKan00,Val03b,Val04a}. It also motivates
and complements some recent results given in \cite{Fri03c}.

The article is structured as follows: sections 2 and 3 provide a brief
discussion of the so-called regular finite initial value problem near
spatial infinity. This discussion is by no means intended to be
comprehensive. This material is based on reference \cite{Fri98a},
to which the reader is remitted for full details. The references
\cite{Fri03c,Fri03a,FriKan00,Val04a,Val03a} may also prove useful. Section
4 considers the construction of initial data satisfying the regularity
condition (\ref{reg_cond}), while section 5 discusses the
asymptotic expansions near null and spatial infinity obtained from such
data. Sections 6 and 7 are devoted to providing an interpretation to
the results of section 5 by looking at the static solutions and also
to expansions of the Bondi mass near spatial infinity. Finally,
section 8 provides some concluding remarks.

\section{General framework}

In this section we review some ideas on the description
of the region of spacetime ``near null and spatial infinity''. Our
discussion follows closely that given in \cite{Fri98a}, and strives
to keep its notation and nomenclature as much as possible. The
reader is remitted to this reference for the details of the
constructions here considered.

\subsection{Spacetime in a neighbourhood of null and spatial infinities}

Let $(\widetilde{M},\widetilde{g}_{\mu\nu})$ \footnote{Throughout this
work the following conventions will be used. The signature of
spacetime metric is $(+,-,-,-)$, thus space metrics are negative
definite. The indices $\mu$, $\nu$ are spacetime ones with range
$0,\ldots,3$, while $\alpha$, $\beta$ are spatial indices with range
$1,\dots,3$. Given a 3 dimensional orthonormal frame $e_{(i)}$, the
indices $i$, $j$ denote components with respect to such frame,
$i,j=1,2,3$. Finally $a,b,\ldots$ and their primed counterparts
$a',b',\ldots$ are spinorial indices taking the values $0,1$.} be a
vacuum spacetime arising as the development, via the Einstein field
equations, of some asymptotically Euclidean initial data
$(\widetilde{S},\widetilde{h}_{\alpha\beta},\widetilde{\chi}_{\alpha\beta})$
with vanishing second fundamental form, $\widetilde{\chi}_{\alpha\beta}=0$ 
---i.e. time symmetric initial data, so that the resulting
development has time reflexion symmetry. The metric
$\widetilde{h}_{\alpha\beta}$ of the initial hypersurface
$\widetilde{S}$ will be taken to be negative definite. For simplicity
and definiteness, it will be assumed that $\widetilde{S}$ contains
only one asymptotically flat region. It will be further assumed that
in this single asymptotically Euclidean region coordinates $y^\alpha$ can be
introduced such that
\[
\widetilde{h}_{\alpha\beta}=-\left(1+\frac{2m}{|y|}\right)\delta_{\alpha\beta}+\O\left(\frac{1}{|y|^2}\right),
\]
as $|y|\rightarrow\infty$ and $m$ is the ADM mass of the initial hypersurface.

In order to discuss the behaviour of the development of the initial data sets  in the
asymptotic region we will make use of the so-called conformal picture.
Accordingly, we assume that there is a 3-dimensional, orientable,
smooth compact manifold $(S,h)$ with a point $i\in S$, and a diffeomorphism
$\Phi$ of $S\backslash\{i\}$ onto $\widetilde{S}$, and a conformal factor $\Omega$
analytic on $S$ with the properties
\begin{eqnarray*}
&& \Omega=0,\mbox{ } d\Omega=0, \mbox{ Hess}(\Omega)=-2h \mbox{ at }i, \\
&& \Omega>0 \mbox{ on } S\backslash\{i\}, \\
&& h=\Omega^2\Phi_*\widetilde{h} \mbox{ on } S\backslash\{i\}.
\end{eqnarray*}
It can therefore be checked that under the above conditions
\[
\Phi^{-1}(y^\alpha)\rightarrow i \mbox{ as } |y|\rightarrow \infty,
\]
so that the point $i$ can be rightfully identified with the infinity
of the initial hypersurface $\widetilde{S}$. It will be assumed that
the conformal factor $\Omega$ is obtained as a result of solving the
Einstein constraint equations. For conceptual reasons we shall
distinguish the point at infinity $i$ of the initial hypersurface
$\widetilde{S}$ from the point $i^0$ corresponding to spatial infinity
in Penrose's framework.

Now, consider an open ball, $B_a(i)\subset S$ of radius $a$ centered on
$i$.  The radius $a$ is chosen to be so that the open ball is
geodesically convex. Let $\rho$ be the geodesic distance along
geodesics emanating from $i$. Furthermore, let $\widetilde{N}$ be the
domain of influence of the set $\Phi(B_a(i)\backslash\{i\})$. On
intuitive grounds we will refer to $\widetilde{N}$ as being the region
of spacetime ``close to null and spatial infinities".
   
It will be convenient for our discussion to blow up the point $i$
representing the infinity of $\widetilde{S}$ to a 2-dimensional
sphere. As we are going to make use of a spinorial formalism, the blow
up of $i$ is technically achieved by resorting to the bundle $SU(S)$
of normalised spin frames over $S$ with structure group
$SU(2,\Complex)$, and projection $\pi$ onto $S$. The details of the
blow up will not be given here. The reader is therefore remitted to
\cite{Fri98a}.

It has been shown in \cite{Fri95} that the use of a 
gauge based on the properties of certain curves known as \emph{conformal
geodesics} renders a conformal factor $\Theta$ that can be read
off directly from the initial data, and thus providing an \emph{a
priori} knowledge of the location of the conformal boundary. The
conformal factor is given by
\begin{equation}
\label{st_cf}
\Theta=\kappa^{-1}\Omega\left(1-\tau^2\frac{\kappa^2}{\omega^2}\right),
\end{equation}
where 
\begin{equation}
\omega=\frac{2\Omega}{\sqrt{|h(d\Omega,d\Omega)|}},
\end{equation}
and $\kappa$ is an arbitrary function on $S$ such that
$\kappa=\rho\kappa'$, with $\kappa'$ analytic and $\kappa'(i)=1$.

Now, let us consider the submanifold $C_{a,\kappa}$ of
$\Real\times\Real\times SU(2,\Complex)$ given by
\[
C_{a,\kappa}=\bigg\{(\tau,\rho,t)\in \Real\times \Real \times SU(2,\Complex)\; |\; 0\leq \rho < a, \;\; -\frac{\omega}{\kappa} \leq \tau \leq \frac{\omega}{\kappa} \bigg\}.
\]
Then it can be shown that there is a projection $\pi$ of $C_{a,\kappa}$ onto $\widetilde{N}$ that can be factored as
\[
C_{a,\kappa} \stackrel{\pi_1}{\longrightarrow} C_{a,\kappa}/U(1) \stackrel{\pi_2}{\longrightarrow} \widetilde{N},\]

The boundary of $C_{a,\kappa}$ is made of the following submanifolds
\begin{subequations}
\begin{eqnarray}
&& I=\{(\tau,\rho,t)\in C_{a,\kappa} \; |\; \rho=0, \;\;|\tau|<1 \}, \\
&& I^\pm=\{(\tau,\rho,t)\in C_{a,\kappa} \;  |\;\rho=0, \;\;\tau=\pm 1\}, \label{degenerate_sets}\\
&& \scri^\pm=\left\{(\tau,\rho,t)\in C_{a,\kappa} \; |\; \rho>0 ,\;\;
\tau=\pm\frac{\omega}{\kappa}\right\}.
\end{eqnarray}   
\end{subequations}
Calculations in the paradigmatic case of the Minkowski solution show
that spacelike geodesics in the physical (unrescaled) spacetime
escaping to infinity, map to spatial curves in $C_{a,\kappa}$
terminating at $I$. Similarly null geodesics in the physical spacetime
map to geodesics in $C_{a,\kappa}$ with end points at $\scri^\pm$
---for more details see \cite{Val03a}. This justifies the names of
\emph{cylinder at spatial infinity} for $I$, and of \emph{null
infinity} for $\scri^\pm$. The sets $I^\pm$ will be considered as
neither belonging to $I$ nor to $\scri^\pm$. They will be refered to as
the sets where spatial infinity ``touches'' null infinity.  The set
\[
\Big\{(\tau,\rho,t)\in C_{a,\kappa} \; | \; \tau=0\Big\}
\]
can be, in a natural way with the (compactified) initial hypersurface
$S$ once the point at infinity has been blown up.

\begin{figure}[t]
\centering
\includegraphics[width=.4\textwidth]{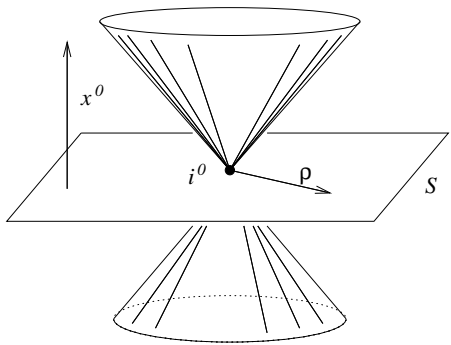} 
\caption[]{The region of (unphysical) spacetime near spatial and null
infinities in the standard representation of spatial infinity as a
point. Note that spacetime is the region lying outside the cone
passing through $i^0$.}
\label{old:fig}
\end{figure}

 The function $\kappa$ appearing in the above expressions reflects the
remaining bit of conformal freedom available in the framework. It can
be used to ``fix'' the shape of null infinity. Choosing $\kappa=1$
brings us back to the conformal representation where spatial infinity
corresponds to a point ---see \cite{Fri98a,Val03a} and the figure
\ref{old:fig}. More interesting for us is to consider a function
$\kappa$ of the form $\kappa=\rho\kappa'$ where $\kappa'$ is an
arbitrary positive analytic function on $B_a(i)$ such that
$\kappa'(i)=1$. Here, we will make use of two such choices of
$\kappa$. Namely,
\[
\kappa_1 =\rho, \qquad \kappa_2=\omega.
\]

\begin{figure}[t]
\centering
\begin{tabular}{cc}
\includegraphics[width=.4\textwidth]{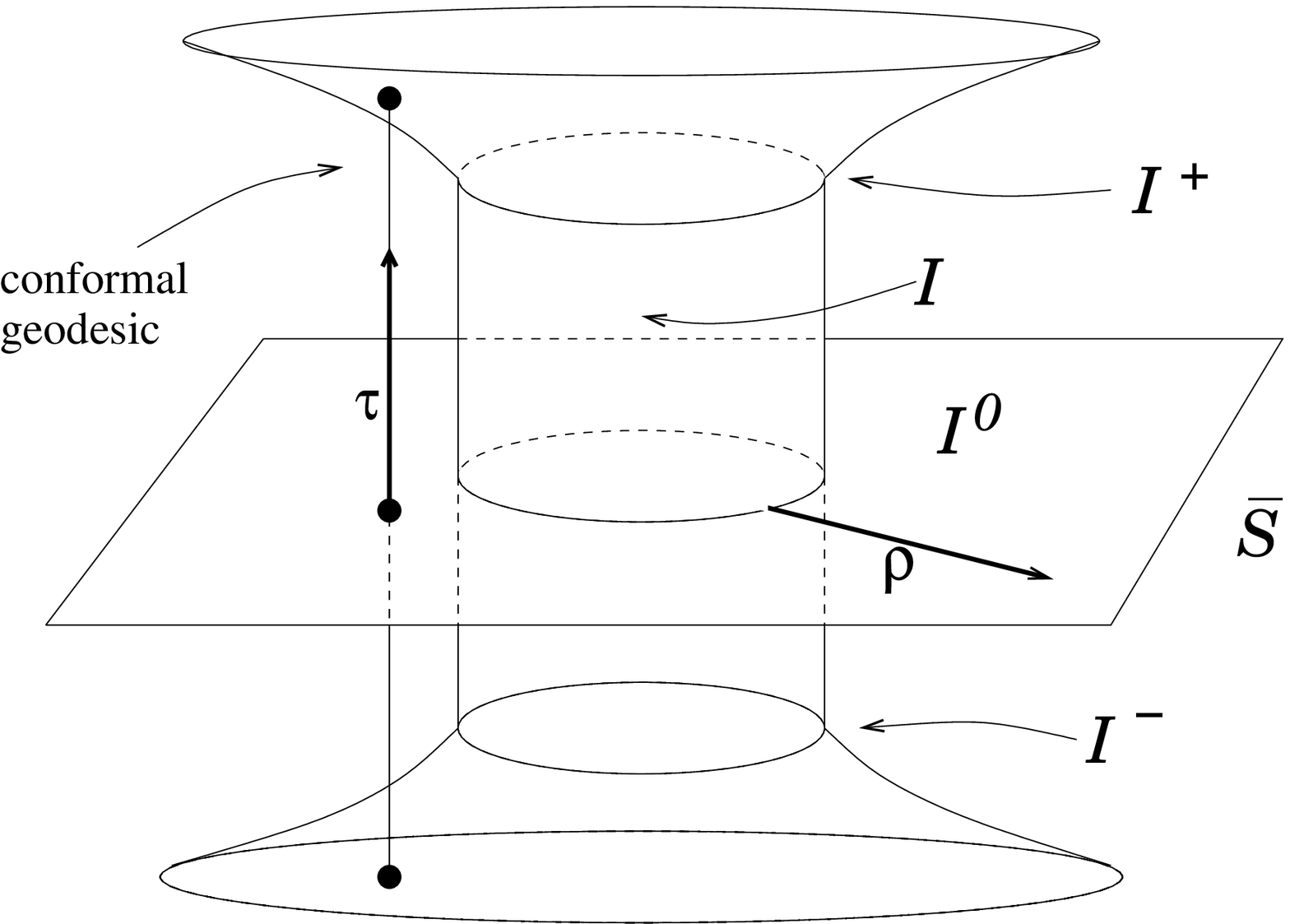} &
\includegraphics[width=.4\textwidth]{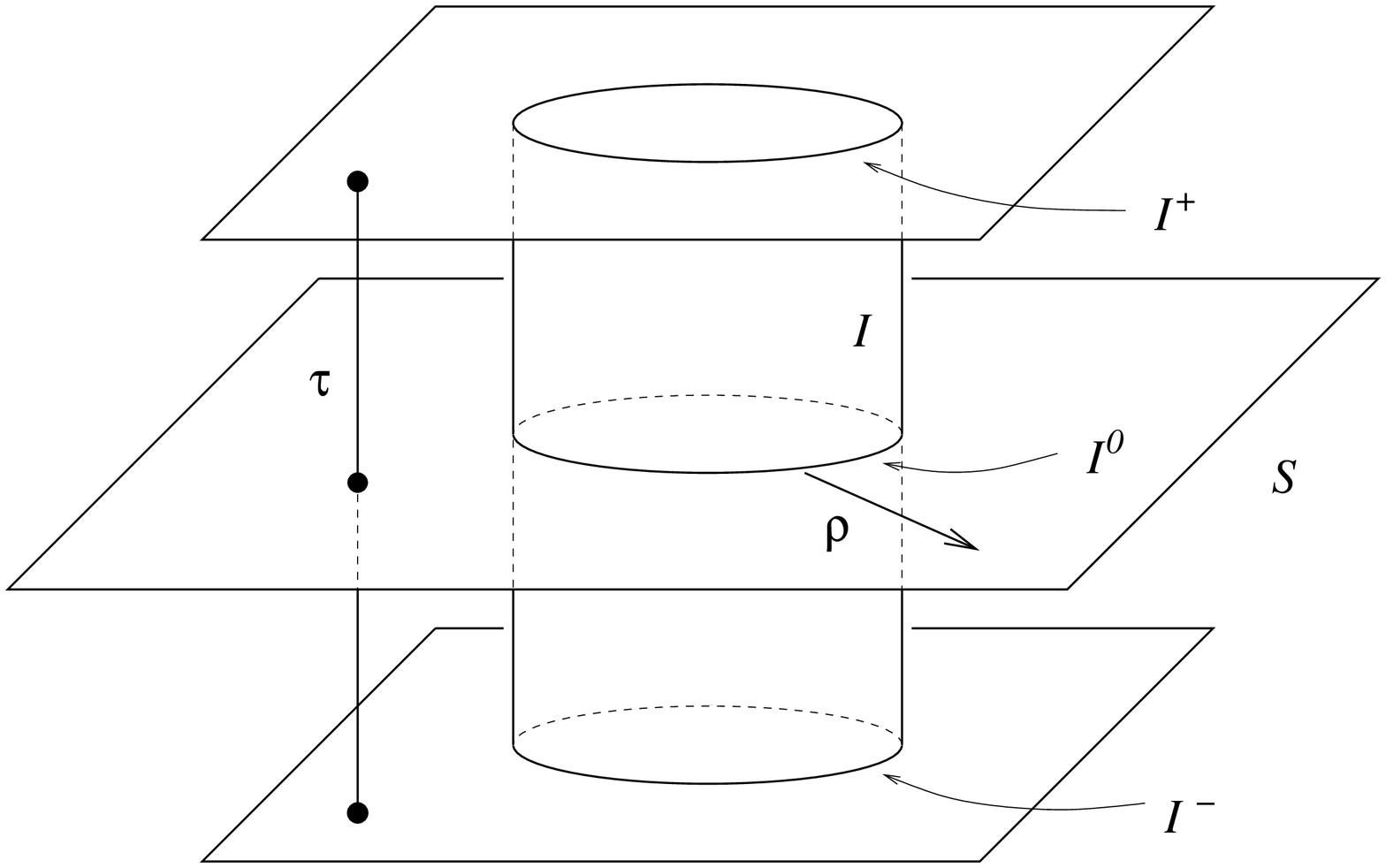}
\end{tabular}
\put(-25,50){$\scri^+$}
\put(-25,-30){$\scri^-$}
\put(-200,60){$\scri^+$}
\put(-200,-50){$\scri^-$}
\caption[]{The representation of spatial infinity using the gauge
  based on the properties of conformal geodesics. The figure to the
  left corresponds to the choice $\kappa=\rho$; the one to the right
  corresponds to $\kappa=\omega$ so that the loci of null infinity are
  hypersurfaces $\tau=\pm 1$.}
\label{new:fig}
\end{figure}

The advantage of the first one renders particularly simple analytic
expression. The second choice has the peculiarity of yielding an
extremely simple formula for the location of null infinity: $\tau=\pm
1$ ---see figure \ref{new:fig}. We will make particular use of this last
choice when calculating the Bondi mass near spatial infinity.

\section{The regular finite initial value problem near spatial infinity}
The finite representation of the region of spacetime near spatial and
null infinity briefly surveyed in the previous section allows the
formulation of an initial value problem near spacelike infinity such
that: the data and equations are regular; the location and structure
of null and spatial is known a priori; and the setting depending on
general properties of conformal structures. We refer to this problem
as to the \emph{regular finite initial value problem near spatial
infinity}.  We now proceed to review some of its features which
are relevant for our analysis.

\subsection{The propagation equations} 
\label{subsect:propeqns}

The propagation equations used by Friedrich in his analysis of
the regular finite initial value problem near spatial infinity on
\cite{Fri98a} are given in a space spinor formalism. This formalism can
be thought as the spinorial analogous of the $1+3$ tensorial
decompositions. The introduction of such a formalism allows us to work
with spinorial quantities having only unprimmed indices. The
(timelike) vectorial field on which this decomposition is performed is
tangent to certain conformal geodesics. For further details on the
space spinor decomposition and the derivation of the propagation
equations, the reader is remitted to \cite{Som80,Fri95,Fri98a}.

The equations given in \cite{Fri98a} imply propagation equations for:

\begin{itemize}
\item[(i)] the components of the frame $c^\mu_{ab}$, $\mu=0,1,\pm$;
the connection coefficients $\Gamma_{abcd}$ which we
decompose as follows
\[
\Gamma_{abcd}=\frac{1}{\sqrt{2}}\left(\xi_{abcd}-\chi_{(ab)cd}\right)-\frac{1}{2}\epsilon_{ab}f_{cd};
\]

\item[(ii)] the Ricci spinor $\Theta_{abcd}$ which also by convenience is to be decomposed as follows
\[
\Theta_{abcd}=\Theta_{(ab)cd}-\frac{1}{2}\epsilon_{ab}\Theta_{g\phantom{g}cd}^{\phantom{g}g},
\]
with $\Theta_{(ab)cd}=\Theta_{(ab)(cd)}$ as $\Theta_{abc}^{\phantom{abc}c}=0$; 

\item[(iii)] the components of the Weyl tensor $\phi_{abcd}=\phi_{(abcd)}$, more usually given in terms of the quantities
\[
\phi_j=\phi_{(abcd)_j}, \qquad j=0,\dots,4,
\]
where subindex $j$ in $(abcd)_j$ indicates that after
symmetrisation, $j$ indices are to be set equal to $1$. 
\end{itemize}

The propagation equations group naturally in two sets: the 
equations for what will be known as the $v$-quantities
$v=\left(c^\mu_{ab},\xi_{abcd},f_{ab},\chi_{(ab)cd},\Theta_{(ab)cd},\Theta_{g\phantom{g}cd}^{\phantom{g}g}
\right)$, $\mu=0,1,\pm$, which are of the form
\begin{equation}
\partial_\tau v=Kv+Q(v,v)+L\phi, \label{v_propagation}
\end{equation}
where $K$ and $Q$ are respectively linear and quadratic functions with
constant coefficients, and $L$ denotes a linear function depending on
the coordinates via the functions $\Theta$, $\partial_\tau \Theta$ and
a 1-form $d_{ab}$. The linear function $L$ is such that $L|_I=0$. For
a detailed listing of the equations, see \cite{Fri98a,Val04a}. Note
that the system (\ref{v_propagation}) is essentially a systems of
ordinary differential equations for the components of the vector $v$.

The second set of equations is, arguably, the most important part of
the propagation equations and corresponds to the evolution equations
for the spinor $\phi_{abcd}$ derived from the Bianchi identities, the
\emph{Bianchi propagation equations} which are of the form
\begin{equation}
\label{bianchi_propagation}
\left( \sqrt{2}E+A^{ab}c^0_{ab}\right)\partial_\tau\phi+A^{ab}c^1_{ab}\partial_\rho\phi+A^{ab}c^C_{ab}\partial_C\phi=B(\Gamma_{abcd})\phi, 
\end{equation}
where $\phi=(\phi_0,\phi_1,\phi_2,\phi_3,\phi_4)$, $\partial_C=X_\pm$
---see section 4---, $E$ denotes the $(5\times 5)$ unit matrix,
$A^{ab}c^\mu_{ab}$ are $(5\times 5)$ matrices depending on the
coordinates, and $B(\Gamma_{abcd})$ is a linear matrix valued function
of the connection. One has that
$(A^{ab}c^1_{ab})|_I=0$.

To the Bianchi propagation equations we add yet another set of three
equations, also implied by the Bianchi identities which we refer to as 
the \emph{Bianchi constraint equations}. These are of the form
\begin{equation}
\label{bianchi_constraint}
F^{ab}c_{ab}^0\partial_\tau\phi +F^{ab}c_{ab}^1\partial_\rho\phi+
F^{ab}c_{ab}^C\partial_C=H(\Gamma_{abcd})\phi,
\end{equation} 
where $F^{ab}c^\mu_{ab}$ denote $(3\times 5)$ matrices, and
$H(\Gamma_{abcd})$ is another matrix valued function of the
connection. It turns out that $(F^{ab}c_{ab}^1)|_{\tau=0}=0$ which
justifies the ``constraint'' label attached to these equations.

The system of equations (\ref{v_propagation})-(\ref{bianchi_propagation})
is to be supplemented by initial data which can be constructed as
follows
\begin{subequations}
\begin{eqnarray}
&& \Theta_{abcd}=-\frac{\kappa^2}{\Omega}
D_{(ab}D_{cd)}\Omega+\frac{1}{12}\kappa r h_{abcd}, \label{initial1}\\
&& \phi_{abcd}=\frac{\kappa^3}{\Omega^2}\left(D_{(ab}D_{cd)}\Omega+\Omega s_{abcd}\right), \label{initial2}\\
&& c^0_{ab}=0, \quad c^1_{ab}=\kappa x_{ab},  \label{initial3}\\
&& c^+=\kappa\left(\frac{1}{\rho}z_{ab}+\check{c}^+_{ab}\right), \quad
c^-=\kappa \left(\frac{1}{\rho}y_{ab}+\check{c}^-_{ab}\right),
 \label{initial4}\\
&& \xi_{abcd}=\sqrt{2}\left\{ \kappa\left(\frac{1}{2\rho}(\epsilon_{ac}x_{bd}+\epsilon_{bd}x_{ac}) + \check{\gamma}_{abcd}\right)
  -\frac{1}{2\kappa}(\epsilon_{ac}D_{bd}\kappa+\epsilon_{bd}D_{ac}\kappa)\right\}, \label{initial5}\\
&& \chi_{(ab)cd}=0, \quad f_{ab}=D_{ab}\kappa. \label{initial6}
\end{eqnarray} 
\end{subequations}
Here, $\Omega$ denotes the conformal factor of the initial
hypersurface $S$, $s_{abcd}$ the spinorial representation of the
symmetric trace free part of the Ricci tensor at the initial
hypersurface, $r$ is its Ricci scalar, $D_{ab}$ denotes the covariant
derivative on $S$, and $\check{c}^\pm_{ab}$ and $\check{\gamma}_{abcd}$ are
the regular parts of the frame and connection on $S$,
respectively. The spinors $x_{ab}$, $y_{ab}$, $z_{ab}$ and $h_{abcbd}$
are defined in section 4.  Recalling that, $\kappa=\kappa'\rho$ with
$\kappa'(i)=1$, then quantities defined by equations
(\ref{initial1})-(\ref{initial6}) are regular for $\rho=0$ if
$\Omega$, $s_{abcd}$ and $r$ arise from time symmetric initial data
with a analytic conformal completion.

\subsection{Transport equations.}
 For $p=0,1,2,\ldots$ let $v^{(p)}$ denote the restriction of
$\partial^p_\rho v$ to $I$, that is, $v^{(p)}=\partial_\rho^p
v|_{\rho=0}$. Similarly, we write, $\phi^{(p)}=\partial_\rho^p
\phi|_{\rho=0}$. For a given integer $p\geq 0$ we will refer to the
set of functions $v^{(p')}$, $0\leq p' \leq p$ as the \emph{s-jet of
order $p$ on $I$}, and denote it by $J^{(p)}_I(v)$. Similar thing with
$J^{(p)}_I(\phi)$. If the solutions $v^{(p)}$ ---or $\phi^{(p)}$---
are evaluated at the intersection of the cylinder $I$ with the initial
hypersurface $S$, then we will talk of a \emph{d-jet of order $p$ on
$I^0$}, and denote it by $J^{(p)}_{I^0}(v)$ ---or
$J^{(p)}_{I^0}(\phi)$ respectively. The knowledge of the s-jets,
$J^{(p)}_I(v)$ and $J^{(p)}_I(\phi)$ allows to construct the following
Taylor polynomial-like expressions for the vectors $v$ and $\phi$:
\[
\sum^p_{k=0}\frac{1}{k!} v^{(k)}\rho^k, \qquad \sum^p_{k=0}\frac{1}{k!} \phi^{(k)}\rho^k.
\]
We will refer to the latter as \emph{the order $p$ normal expansions of
$v$ and $\phi$}. These expansions are to be understood in the sense of
truncated series. Our current analytic understanding of the of the
propagation equations does not allow us to obtain estimates of the
remainders of these expressions. However, it is noted that in the case
of linear gravity it has been possible to construct some
(non-standard) estimates \cite{Fri03b}.

The analysis carried out in \cite{Val03a} and continued in the present
article makes use of the fact that the cylinder at spatial infinity
$I$ is a \emph{total characteristic} of the propagation equations
(\ref{v_propagation}), (\ref{bianchi_propagation}) and
(\ref{bianchi_constraint}). This is closely related to the fact that
$(A^{ab}c^1_{ab})|_I=0$ ---see \cite{Fri98a,Val04a} for the
details. This general feature of our framework allows us to calculate
the vectors $v^{(p)}$ and $\phi^{(p)}$ from a knowledge of the jets
$J^{(p-1)}_I(v)$ and $J^{(p-1)}_I(\phi)$. The equations governing the
vectors $v^{(p)}$ and $\phi^{(p)}$ are known as the \emph{$p$th order
transport equations}. These can be written
as
\begin{eqnarray}
&&\partial_\tau v^{(p)} = Kv^{(p)}
+Q(v^{(0)},v^{(p)})+Q(v^{(p)},v^{(0)}) \nonumber \\
&&\phantom{XXXXXXXX}+\sum_{j=1}^{p-1}\left(
  Q(v^{(j)},v^{(p-j)})+ L^{(j)}\phi^{(p-j)}\right) +
L^{(p)}\phi^{(0)}. \label{v_transport}
\end{eqnarray} 

From the Bianchi propagation equations (\ref{bianchi_propagation}) one gets
\begin{eqnarray}
&&\left( \sqrt{2}E +A^{ab}(c^0_{ab})^{(0)}\right)\partial_\tau\phi^{(p)} +
A^{ab}(c_{ab}^C)^{(0)}\partial_C\phi^{(p)}= B(\Gamma^{(0)}_{abcd})\phi^{(p)}
\nonumber \\
&&\phantom{XXXXXXXX}+\sum_{j=1}^p
\left(\begin{array}{c} p \\ j
  \end{array}\right)\left(B(\Gamma_{abcd}^{(j)})\phi^{(p-j)}-A^{ab}(c^\mu_{ab})^{(j)}\partial_\mu
 \phi^{(p-j)}\right). \label{b_transport}
\end{eqnarray}
Similarly, from the Bianchi constraint equations
(\ref{bianchi_constraint}) one obtains
\begin{eqnarray}
&&F^{ab}(c^0_{ab})^{(0)}\partial_\tau\phi^{(p)}+F^{ab}(c^C_{ab})^{(0)}\partial_C\phi^{(p)}=H(\Gamma^{(0)}_{abcd})\phi^{(p)}
  \nonumber \\
&&\phantom{XXXXXXXX}+\sum_{j=1}^p
\left(\begin{array}{c} p \\ j
  \end{array}\right)
 \left(
   H(\Gamma_{abcd}^{(j)})\phi^{(p-j)}-F^{ab}(c^\mu_{ab})^{(j)}\partial_\mu\phi^{(p-j)}\right). \label{c_transport}
\end{eqnarray}

Diverse properties of the transport equations and how these
can be solved are found in \cite{Fri98a,FriKan00,Val03a}. In
particular in \cite{Val03a} it has been briefly discussed how it is
possible to solve these equations by means of computer algebra
methods.

\section{Constructing initial fulfilling the regularity condition}
As seen in section \ref{subsect:propeqns}, the initial data for the
propagation equations (\ref{v_propagation}), (\ref{bianchi_propagation})
---and consequently also for the transport equations
(\ref{v_transport})-(\ref{b_transport})--- can be constructed from a
knowledge of the conformal factor $\Omega$, the symmetric and tracefree Ricci
spinor $s_{abcd}$, the Ricci scalar $r$, and the regular part of the
frame coefficients and connection, $\check{c}^1_{ab}$,
$\check{c}^\pm_{ab}$, $\check{\gamma}_{abcd}$ ---see equations
(\ref{initial1})-(\ref{initial6}). Under our assumption of time
symmetry, the equation determining $\Omega$ arises from the
Hamiltonian constraint by making the so-called conformal Ansatz.  The
resulting equation is the \emph{Licnerowicz equation}
\begin{equation}
\label{licnerowicz_equation}
\left(\Delta_h -\frac{1}{8}r\right)\left(\Omega^{-1/2}\right)=4\pi \delta_i,
\end{equation}
where $\delta_i$ is the Dirac-delta function with support on $i$, and $\Delta_h$ is the Laplacian with respect to the metric $h$. It
is customary to use the following parametrisation for the conformal
factor $\Omega$ in a neighbourhood of infinity, $B_a(i)$:
\[
\Omega=\frac{\rho^2}{(U+\rho W)^2},
\]
where $U$ contains information about the local geometry around $i$, whereas the function $W$ contains information of global nature ---e.g. about the ADM mass and higher order multipoles. One has:
\begin{eqnarray}
&& \left(\Delta_h -\frac{1}{8}r\right)\left(\frac{U}{\rho^2}\right)=4\pi \delta_i, \label{yamabe1}\\
&& \left(\Delta_h -\frac{1}{8}r\right)W=0 \label{yamabe2},
\end{eqnarray}
near $i$. Furthermore,
\[
U(i)=1, \qquad W(i)=\frac{m}{2}.
\]
 Because of the nature of our analysis, we will be just interested in
 constructing jets $J^{(p)}_{I^0}(U)$ and $J^{(p)}_{I^0}(W)$ for a
 certain non-negative integer $p$ consistent with equations
 (\ref{yamabe1}) and (\ref{yamabe2}).

\subsection{Freely specifiable data at each order}

In the conformal method to solve the constraint equations, the freely
specifiable data under the assumption of time symmetry is given in
terms of the conformally rescaled metric $h_{\alpha\beta}$. Due to 
our coordinate choice ---$\rho$ is a geodesic distance--- it has the
form
\[
h_{\alpha\beta}=\left(
\begin{array}{ccc}
1 & 0 & 0 \\
0 & h_{22} & h_{23} \\
0 & h_{23} & h_{33}
\end{array}
\right).
\]
Now, consider an orthonormal frame $e_{(i)}$
---$h_{\alpha\beta}e^\alpha_{(i)}e^\beta_{(j)}=-\delta_{(i)(j)}$. Associated
to the frame $e_{(i)}$ there is a certain spinorial field
$c_{ab}$. The correspondence between the two is given by the 
spatial Infeld-van der Waerden symbols $\sigma^{(i)}_{ab}$, in the form
\[
c_{ab}=\sigma^{(i)}_{ab}e_{(i)}.
\] 
Associated to the fields $s_{abcd}$, $r$, $c_{ab}$ on $S$ there are
corresponding ---$U(1)$ invariant--- lifts to the set
$\Sigma_a=\{(\tau,\rho,t)\in C_{a,\kappa} | \tau=0\}$ which in an
abuse of notation we denote by the same symbols. The spinorial field
$c_{ab}$ on the initial hypersurface can be written as
\[
c_{ab}=c_{(ab)}=x_{ab}\partial_\rho + \left( \frac{1}{\rho}z_{ab}+ \check{c}^+_{ab}\right)X_+ + \left( \frac{1}{\rho}y_{ab} +\check{c}^-_{ab}\right)X_-,
\]
where $X_\pm$ are differential operators on $SU(2,\Complex)$ related to the
$\eth$, $\overline{\eth}$ operators of the NP formalism ---see
\cite{Fri98a} for full details. The coefficients $\check{c}_{ab}^\pm$, \emph{the regular parts of the frame}, satisfy
\[
\check{c}^\pm_{ab}=\O(\rho), \qquad \check{c}^\pm_{01}=0.
\] 
The \emph{elementary spinors} $x_{ab}$, $y_{ab}$, $z_{ab}$ are defined as follows
\[
x_{ab}=\sqrt{2}\epsilon_{(a}^{\phantom{a}0}\epsilon_{b)}^{\phantom{b)}1},
\quad y_{ab}=-\frac{1}{\sqrt{2}} \epsilon_{a}^{\phantom{a}1}
\epsilon_{b}^{\phantom{b}1}, \quad   z_{ab}=\frac{1}{\sqrt{2}} \epsilon_{a}^{\phantom{a}0}
\epsilon_{b}^{\phantom{b}0}.
\]
We shall also use the spinor
\[
h_{abcd}=-\epsilon_{a(c}\epsilon_{d)b},
\]
corresponding to the components of the initial metric
$h_{\alpha\beta}$ with respect to the spin basis we are using. 

The field $c_{ab}$ satisfies the following reality conditions
\[
\check{c}^-_{00}=\overline{\check{c}}^+_{11}, \qquad
\check{c}^-_{11}=\overline{\check{c}}^+_{00}. 
\] 
We shall denote the connection associated with the frame $c_{ab}$ by
$\gamma_{abcd}$. Given a spinorial field $\mu_{cd}$, its covarinat
derivative is given by
\[
D_{ab}\mu_{cd}=c_{ab}(\mu_{cd})-\gamma_{ab\phantom{e}c}^e\mu_{ed}-\gamma_{ab\phantom{e}d}^e\mu_{ce}.
\]
We decompose $\gamma_{abcd}$ into a singular and
a regular parts ---in concordance with equation (\ref{initial5})--- as follows
\[
\gamma_{abcd}=\frac{1}{2\rho}(\epsilon_{ac}\epsilon_{bd}+\epsilon_{bd}\epsilon_{ac})+\check{\gamma}_{abcd},
\]
where $\check{\gamma}_{abcd}=\O(\rho)$. The frame coefficients and
the connection are related via commutator equations entailing
\begin{subequations}
\begin{eqnarray}
&& \frac{1}{\sqrt{2}}\partial_\rho(\rho\check{c}^+_{aa})=\check{\gamma}_{aa00}(\rho\check{c}^+_{11})-\check{\gamma}_{aa11}(\rho\check{c}^+_{00}) -\frac{1}{\sqrt{2}}\check{\gamma}_{aa11}, \label{commutator_1}\\
&& \frac{1}{\sqrt{2}}\partial_\rho(\rho\check{c}^-_{aa})=\check{\gamma}_{aa00}(\rho\check{c}^-_{11})-\check{\gamma}_{aa11}(\rho\check{c}^-_{00}) -\frac{1}{\sqrt{2}}\check{\gamma}_{aa00} \label{commutator_2}.
\end{eqnarray}
\end{subequations}
Now, the connection coefficients $\check{\gamma}_{abcd}$ satisfy
\[
\check{\gamma}_{01cd}=0, \qquad \check{\gamma}_{1100}=-\check{\gamma}_{0011}, \qquad \check{\gamma}_{abcd}=\check{\gamma}_{abdc},
\]
so there are only 5 independent connection coefficients:
$\check{\gamma}_{0000}$, $\check{\gamma}_{0001}$,
$\check{\gamma}_{0011}$, $\check{\gamma}_{1101}$, and
$\check{\gamma}_{1111}$ modulo the reality conditions
\[
\overline{\check{\gamma}}_{1101}=\overline{\check{\gamma}}_{1110}=\check{\gamma}_{0010}=\check{\gamma}_{0001}.
\]
The commutator equations (\ref{commutator_1})-(\ref{commutator_2})
allow to fully determine the frame coefficients $\check{c}^\pm_{ab}$
in terms of the connection coefficients $\check{\gamma}_{abcd}$. The
connection coefficients and the curvature are related via the
structure equations of the initial hypersurface. In our case the
non-trivial ones can be written as
\begin{subequations}
\begin{eqnarray}
&&\frac{1}{\sqrt{2}} \partial_\rho\check{\gamma}_{00ab}+\frac{1}{\rho}\left\{ \check{\gamma}_{0000}z_{ab}-\check{\gamma}_{0011}y_{ab}+\frac{1}{\sqrt{2}}\check{\gamma}_{00ab}\right\} \nonumber \\
&& \phantom{XXXXXXXX}= \check{\gamma}_{0000}\check{\gamma}_{11ab}-\check{\gamma}_{0011}\check{\gamma}_{00ab}-\frac{1}{2}s_{ab00}-\frac{1}{6\sqrt{2}}ry_{ab}, \label{structure_1} \\
&&\frac{1}{\sqrt{2}} \partial_\rho\check{\gamma}_{11ab}+\frac{1}{\rho}\left\{ \check{\gamma}_{1100}z_{ab}-\check{\gamma}_{1111}y_{ab}+\frac{1}{\sqrt{2}}\check{\gamma}_{11ab}\right\}  \nonumber\\
&& \phantom{XXXXXXXX}= \check{\gamma}_{1100}\check{\gamma}_{11ab}-\check{\gamma}_{1111}\check{\gamma}_{00ab}-\frac{1}{2}s_{ab11}-\frac{1}{6\sqrt{2}}rz_{ab} \label{structure_2}.
\end{eqnarray}
\end{subequations}
The latter, in turn, allow to write the connection coefficients
$\check{\gamma}_{abcd}$ in terms of the traceless Ricci spinor
$s_{abcd}$ and the Ricci scalar $r$. Consequently, also the frame
coefficients $c^\pm_{ab}$ can be fully written in terms of
$s_{abcd}$ and $r$. The components of $s_{abcd}$ and $r$ are, however,
not independent. They satisfy the 3-dimensional Bianchi identity
\begin{equation}
\label{bianchi_3d}
D^{ab}s_{abcd}=\frac{1}{6}D_{cd}r.
\end{equation}
 Because of its symmetries, the traceless Ricci spinor can be decomposed in
 terms of elementary spinors as
\[
s_{abcd}=s_0\epsilon^0_{abcd}+s_1\epsilon^1_{abcd}+s_2\epsilon^2_{abcd}+s_3\epsilon^3_{abcd}+s_4\epsilon^2_{abcd},
\]  
where the coefficients $s_j$ are still subject to the reality conditions
\[
s_4=\overline{s}_0, \;\; s_3=-\overline{s}_1, \;\; s_2=\overline{s}_2,
\]
and
\[
\epsilon^j_{abcd}=\epsilon_{(a}^{\phantom{(a}(e}\epsilon_{b}^{\phantom{b}f}\epsilon_{c}^{\phantom{c}g}\epsilon_{d)}^{\phantom{d)}h)_j}.
\]

 Details on decomposition of spinors in terms of irreducible terms and
 on the reality conditions can be found in \cite{Fri98a,FriKan00}.  

In the sequel, it will be convenient to regard the Bianchi identity
 (\ref{bianchi_3d}) as providing 3 equations from which the components
 $s_1$, $s_2$ and $s_3$ can be determined in terms of the remaining
 two components $s_0$, $s_4$ and the Ricci scalar $r$.

The spinorial field $s_{abcd}$ and the scalar $r$ are functions on
 $\Real\times SU(2,\Complex)$. Therefore, they admit a decomposition
 in terms of certain functions, $T_{m\phantom{j}k}^{\phantom{m}j}$,
 $m=0, 1,2,\ldots$, $j,k=0,\ldots,m$, associated with unitary
 representations of $SU(2,\Complex)$ ---the index $m$ being the label
 of the representation. The functions
 $T_{m\phantom{j}k}^{\phantom{m}j}$ are closely related to
 spin-weighted spherical harmonics via the correspondence ---see
 \cite{FriKan00} for details---:
\[
{}_sY_{nm}\mapsto (-i)^{s+2n-m}\sqrt{\frac{2n+1}{4\pi}}T_{2n\phantom{n-m}n-s}^{\phantom{2n}n-m}.
\]

 Under the assumptions of time symmetry and
 analyticity of the conformal metric $h_{\alpha\beta}$ it can be seen
 ---see again \cite{Fri98a} for the details--- that
\begin{subequations}
\begin{eqnarray}
&& s_j=\sum_{p=1}^\infty \sum_{q=|2-j|}^{p+1} \sum_{k=0}^{2q}\frac{1}{p!} s_{j,p;2q,k}T_{2q\phantom{k}q-2+j}^{\phantom{2q}k}\rho^p, \label{s_expansion}\\
&& r=\sum_{p=1}^\infty \sum_{q=0}^{p} \sum_{k=0}^{2q} \frac{1}{p!}r_{p;2q,k}T_{2q\phantom{k}q}^{\phantom{2q}k}\rho^p \label{r_expansion},
\end{eqnarray}
\end{subequations}
with $\overline{s}_{j,p;2q,2q-k}=(-1)^{k+q}s_{j,p;2q,k}$ and
$\overline{r}_{p;2q,2q-k}=(-1)^{q+k}r_{p;2q,k}$ as a consequence of
the reality conditions. The commutators
(\ref{commutator_1})-(\ref{commutator_2}) and the structure equations
(\ref{structure_1})-(\ref{structure_2}) then imply that
$\check{c}^\pm_{ab}$ and $\check{\gamma}_{abcd}$ have expansion type
$p$. Direct evaluation ---using {\tt Maple V}--- up to order
$p=6$ shows that,
\[
s_{j,p;2q,k}=s_{j,p;2q,k}( s_{j',p';2q',k'}, r_{p'';2q'',k''}),
\]
where $j=1,2,3$, $p=1,\dots,6$, $q=|2-j|,\ldots,p$, $k=0,\ldots,2q$
for some $j',=0,4$, $1\leq p'\leq p$, $1\leq p'' \leq p$, $|2-j'|\leq
q'\leq q$, $0 \leq q''\leq q$ and $0\leq k' \leq k$, $0\leq k''\leq
k$. Using (\ref{commutator_1})-(\ref{commutator_2}) and
(\ref{structure_1})-(\ref{structure_2}) one can express the
coefficients appearing in the expansions for $\check{\gamma}_{abcd}$,
$\check{c}^\pm_{ab}$ in terms of those appearing in
(\ref{s_expansion}) and (\ref{r_expansion}).

It is worth making a couple of observations. Firstly, the coefficients
in the expansions (\ref{s_expansion}) with $j=0,4$ and in
(\ref{r_expansion}) are not all independent, as these have to be
consistent with the expansions of the constraint equations. Another
observation is that even if the expansions (\ref{s_expansion}) and
(\ref{r_expansion}) are already fully consistent with the constraints,
some of the coefficients happen to be pure gauge. This is because
there is still some conformal freedom left in our construction. How to
identify which coefficients in the expansions are gauge and which not
will also be discussed in the sequel.

\subsection{Solving the Yamabe equation}

The equation for the function $W$, equation (\ref{yamabe2}), is
usually known as the \emph{Yamabe equation}. As a part of our
calculations, it shall be necessary to solve it up to a certain order
$p$. This in turn will provide a knowledge of the jet
$J^{(p)}_I(W)$. If the rescaled metric $h_{\alpha\beta}$ is analytic,
then $W$ is also analytic ---cfr. \cite{DaiFri01}. Thus one can write
\begin{subequations}
\begin{equation}
W=\frac{m}{2}+\sum^\infty_{p=1}\frac{1}{p!}w_{p}\rho^{p},
\end{equation}
 with
\begin{equation}
\label{w_p}
w_{p}=\sum_{q=0}^p \sum_{k=0}^{2q} w_{p;2q,k} T_{2q\phantom{k}q}^{\phantom{2q}k},
\end{equation}
\end{subequations}
where the coefficients $w_{p;2q,k}$ are complex numbers satisfying the
reality condition
$\overline{w}_{p;2q-k}=(-1)^{q+k}w_{p;2q,k}$. Because of the finite
order of our calculations, it will not be necessary to assume that $W$
is analytic. However, we will require that it has an 6th-order Taylor
polynomial in $\rho$ with coefficients of the form given by equation
(\ref{w_p}).

We note that if the initial data is assumed to be conformally flat,
then $W$ is harmonic. This implies that
\[
w_{p}=\sum_{k=0}^{2q} w_{p;2p,k} T_{2p\phantom{k}p}^{\phantom{2p}k}.
\]
If the initial data is not conformally flat, then in the cn-gauge
---see \S\ref{subsect:cngauge}--- the function $W$ is in general only
harmonic up to order $p=3$.

\subsection{The Green function}

The function $U$, solution to equation (\ref{yamabe1}), is also known
as the Green function. It can be determined using what is known as the
Hadamard's parametrix construction. In this construction one assumes
the following Ansatz for the function $U$:
\[
U=\sum^\infty_{p=0} U_p\rho^{2p}.
\]
where $U_p=U_p(\rho,t)$. These can be obtained recursively from
\[
U_0=\exp\left\{ \frac{1}{4}\int_0^{\rho}(\Delta_h\rho^2 +6)\frac{ds}{s}\right\},\qquad
U_{p+1}=-\frac{U_0}{(4p-2)\rho^{p+1}}\int_0^\rho\frac{\Delta_h[U_p]s^p}{U_0}ds.
\]
In particular, if one has conformally flat data then $U=1$. If
$h_{\alpha\beta}$ is analytic near $i$, then the function $U$ is
analytic, and furthermore, the coefficients $U_p$ are also analytic.

\subsection{The cn-gauge}
\label{subsect:cngauge}
The construction of the functions $W$ and $U$ described before does not fix the
functions uniquely. A rescaling of the form
\begin{subequations}
\begin{equation}
\label{rescaling_1}
h\rightarrow h'=\vartheta^4 h, \qquad \Omega\rightarrow \Omega'=\vartheta^2\Omega,
\end{equation}
with a conformal factor $\vartheta$ satisfying $\vartheta(i)=1$, leaves $\widetilde{h}=\Omega^{-2}h$ unchanged, but implies
\begin{equation}
\label{rescaling_2}
U\rightarrow U'=\frac{\rho'}{\rho}\vartheta^{-1}, \qquad  W\rightarrow W'=\vartheta^{-1}W,
\end{equation}
\end{subequations}
where $\rho'$ is the $h'$ geodesic distance along geodesics starting at $i$.
In order to remove this freedom our discussion will done in a certain gauge
known as the \emph{cn-gauge}. Note however, that our whole discussion could
have been carried in another gauge ---for example, the
discussion in \cite{Fri03c} uses a certain gauge for which $W=m/2$.

The \emph{cn-gauge} is defined as follows. Consider the 3-dimensional
conformal geodesic equations
\begin{eqnarray}
&& \dot{x}^\beta\nabla_\beta \dot{x}^\alpha = -2 b_\beta \dot{x}^\beta +
\dot{x}_\beta \dot{x}^\beta b^\alpha,  \label{3d_cg_1}\\
&& \dot{x}^\beta\nabla_\beta b_\alpha = b_\beta \dot{x}^\beta b_\alpha
-\frac{1}{2} b_\beta b^\beta \dot{x}_\alpha + \left(s_{\alpha\beta} +
  \frac{1}{12} r h_{\alpha\beta}\right)\dot{x}^\beta, \label{3d_cg_2}
\end{eqnarray}
where now $x^\alpha(t)$ is a curve on $S$, and $b^\alpha$ an
associated 3-dimensional 1-form. Indices are raised and lowered using
$h_{\alpha\beta}$ and $h^{\alpha\beta}$ respectively. We supplement
the latter equations with the initial conditions
\[
x(0)=i, \qquad \dot{x}_{\beta}\dot{x}^\beta=-1, \qquad b(0)=0.
\]
It is not hard to see that if $a$ is chosen small enough, there exists a
unique solution to these equations on $B_a(i)$. Furthermore, there exist in
$B_a(i)$ a unique rescaling of the form (\ref{rescaling_1})-(\ref{rescaling_2}) such that
\begin{equation}
\label{cn_gauge}
b_{\beta}\dot{x}^\beta=0 \qquad \mbox{ on } B_a(i),
\end{equation}
can always be attained. A metric in the conformal class for which the
condition (\ref{cn_gauge}) is satisfied along the solutions of the
3-dimensional conformal geodesic equations (\ref{3d_cg_1}) and
(\ref{3d_cg_2}) will be said to be in the \emph{cn-gauge}.

It is noticed that for conformally flat data, if $\dot{x}$ is the
tangent to (standard) geodesics starting at $i$ with
$\dot{x}_{\beta}\dot{x}^\beta=-1$, and one requires $b\equiv 0$ in
$B_a(i)$, one is already in the cn-gauge. This is the reason why the
calculations discussed in \cite{DaiVal02,Val03b,Val04a} did not
contemplate this point.

The cn-gauge condition (\ref{cn_gauge}) imposes restrictions on
$s_{\alpha\beta}$ ---and consequently in its spinorial representation
$s_{abcd}$--- and $r$. One can obtain a space-spinor version of the
conformal geodesic equations (\ref{3d_cg_1}) and (\ref{3d_cg_2}) by
contracting with the frame $e_{(i)}$, and then using the spatial Infeld
van der Waerden symbols. These spinorial equations together with the
results of \S4.1  yield after some calculations in {\tt Maple V}
the following result.

\begin{lemma}[the free specifiable data in the cn-gauge]
\label{lemma:freedata_cngauge}
In the cn-gauge one has that
\[
s_{j,1;2q,k}=0, \qquad r_{1;2q,k}=0,
\]
for $j=0,\ldots,4$, $q=0,\ldots,2$, $k=0,\ldots,2q$. Furthermore, 
\[
s_{j,p;2(p+1),k}=0,
\]
for $j=0,\ldots,4$, $p=2,\dots,6$, $k=0,\ldots,2q$. The coefficients of the expansion of the trace free Ricci spinor and the Ricci scalar are of the form
\begin{eqnarray*}
&&s_{j,p;2q,k}=s_{j,p;2q,k}(s_{0,p;2q,k},s_{4,p;2q,k}), \\
&&r_{p;2q,k}=r_{p;2q,k}(s_{0,p;2q,k},s_{4,p;2q,k}),
\end{eqnarray*}
for $j=1,2,3$ and $p=2,\ldots,6$, $q=2,\ldots,p$, $k=0,\ldots,2q$. 
\end{lemma}

It is stressed once more that the results are valid for expansions up
to order $p=6$ ---the order up to which the {\tt Maple V} calculations
have been carried out.

\subsection{Fulfilling the regularity condition}

We begin by recalling the following important result by Friedrich
\cite{Fri98a}.

\begin{theorem}[Friedrich,1998]
The solutions to the transport equations are smooth through $I^\pm$
only if the (regularity) condition
\[
D_{(a_sb_s}\cdots D_{a_1b_1} b_{abcd)}(i)=0, \quad s=0,1,\ldots
\]
is satisfied by the free initial data.
\end{theorem}

As discussed \emph{in extensis} in \cite{Val04a} the great
simplification of studying expansions for conformally flat data
lies in the fact that these satisfy the regularity condition
(\ref{reg_cond}) trivially. In the current analysis the situation is
completely different. Given free data in the cn-gauge ---in the way
discussed in lemma \ref{lemma:freedata_cngauge}--- one still has
specialise to those free specifiable data consistent with the
regularity condition. The question is now, how to implement the
regularity condition? A first ---naive--- approach would be to
calculate directly the spinor $b_{abcd}$ and then its symmetrised
derivatives up to the required order. This approach is computationally
too involved, so we have opted for a different approach. In
\cite{Fri98a} it has been shown that the if the regularity condition
holds up to a certain order, then what is known as the massless part
of the Weyl tensor has a particular form. We shall make use of this
result.

\medskip
The \emph{massless part of the Weyl tensor} is given by
\begin{eqnarray*}
&&\phi'_{abcd}=\frac{1}{\rho^4}\bigg(U^2D_{(ab}D_{cd)}\rho^2 -4UD_{(ab}\rho^2D_{cd)}U  \\
&&\phantom{\phi'_{abcd}=XX}-2\rho^2UD_{(ab}D_{cd)}U +6\rho^2D_{(ab}UD_{cd)}U +\rho^2U^2s_{abcd}\bigg).
\end{eqnarray*}
Let also
\[
\breve{\phi}'_{abcd}=\kappa^3\phi'_{abcd}.
\]
The spinor $\breve{\phi}'_{abcd}$ can be seen to have an expansion of the form
\[
\breve{\phi}'_j=\sum_{p=|2-j|}^\infty \sum_{q=|2-j|}^p \sum_{k=0}^{2q} \frac{1}{p!} \breve{\phi}'_{j,p;2q,k} T_{2q\phantom{k}q-2+j}^{\phantom{2q}k}\rho^p,
\]
where $\breve{\phi}_j'=\breve{\phi}'_{(abcd)_j}$. Using the aforedefined quantities, one has the following lemma.

\begin{lemma}
\label{lemma:equivalence_regcond}
The following two conditions are equivalent:
\begin{itemize}
\item[i)]
\[
D_{(a_qb_q}\cdots D_{a_1b_1}b_{abcd)}(i)=0, \qquad q=0,1,\ldots,s;
\]

\item[ii)]
\[
\breve{\phi}'_{j,p,2p,k}=0, \quad p=0,1,\ldots,s, \quad k=0,\dots,2p, \quad j=0,\ldots,4.
\]

\end{itemize}
\end{lemma}

The proof of lemma \ref{lemma:equivalence_regcond} can be found in
reference \cite{Fri98a}. From lemma \ref{lemma:equivalence_regcond}
direct calculations using {\tt Maple V} lead to the following result.

\begin{lemma}
\label{lemma:cn_regcond}
In the cn-gauge, if the regularity condition (\ref{reg_cond}) holds up
to order $s=4$, then
\[
s_j=\sum_{p=2}^6 \sum_{k=0}^{2p}\frac{1}{p!}s_{j,p;2p,k}T_{2p\phantom{k}p+j-2}^{\phantom{2p}k}\rho^p+\cdots,
\]
with $j=0,\ldots,4$. Furthermore, 
\[
r=\sum_{p=2}^6\sum_{k=0}^{2p}\frac{1}{p!}r_{p;2p,k}T_{2p\phantom{k}p}^{\phantom{2p}k}\rho^p+\cdots.
\]
\end{lemma}

This last result would seem to indicate that $r$ is an harmonic
function. Explicit calculations up to order $p=7$ show that this is
not the case.

\subsection{Initial data for the propagation equations} 
\label{data_for_propagation}
Starting from lemma \ref{lemma:cn_regcond} and taking the discussion
of section 4 backwards, one can calculate d-jets $J^{(6)}_{I^0}(v)$ and
$J^{(6)}_{I^0}(\phi)$ consistent with the regularity condition
(\ref{reg_cond}). We summarise this construction: start with given $s_0$ and
$s_4$ ($s_4=\overline{s}_0$) , components of the spinor $s_{abcd}$
which in the cn-gauge are of the form
\begin{eqnarray}
&&s_0=\sum_{p=2}^6 \sum_{k=0}^{2p}\frac{1}{p!}s_{0,p;2p,k}T_{2p\phantom{k}p-2}^{\phantom{2p}k}\rho^p+\cdots,\label{S0}\\
&&s_4=\sum_{p=2}^6 \sum_{k=0}^{2p}\frac{1}{p!}s_{4,p;2p,k}T_{2p\phantom{k}p+2}^{\phantom{2p}k}\rho^p+\cdots. \label{S4}
\end{eqnarray}
Using the Bianchi identities one can calculate the remaining
components $s_1$, $s_2$, $s_3$, and because of the cn-gauge, also the
Ricci scalar $r$. Now, using the structure equations
(\ref{structure_1})-(\ref{structure_2}) first, and then the commutator
equations (\ref{commutator_1})-(\ref{commutator_2}) one can calculate
the regular part of the connection and the frame,
$\check{\gamma}_{abcd}$, $\check{c}^\pm_{ab}$ up to order $6$
inclusive. Using these, one can construct a function $W$ consistent up
to order $6$ with the help of equation (\ref{yamabe2}). Similarly,
using Hadamard's procedure one can calculate the function $U$ up to
order $7$. The latter requires the evaluation of the steps $0,1,2$ of
Hadamard's recursive procedure.  Because of lemma \ref{lemma:cn_regcond},
such $U$ yields a massless part of the Weyl tensor such that the
regularity condition (\ref{reg_cond}) is satisfied to order
$6$. Knowledge of $W$ to order $6$ and $U$ to order $7$ allows to
calculate the conformal factor $\Omega$ to order $9$ inclusive. This
is exactly what is required in order to calculate the d-jets,
$J^{(6)}_{I^0}(v)$ and $J^{(6)}_{I^0}(\phi)$. These by construction
should be consistent with the regularity condition
(\ref{reg_cond}).
  
A final result ensures that our whole construction of initial data for
the conformal propagation equations is consistent.

\begin{lemma}[fulfillment of the Bianchi constraint equations]
\label{fulfillment}
Assume that the components $s_0$ and $s_4$ of the Ricci spinor are of
the form given by (\ref{S0}) and (\ref{S4}). Further, assume that the
connection and frame coefficients $\check{\gamma}_{abcd}$ and
$\check{c}_{ab}$ have been calculated up to order 6 using the
structure and commutator equations. Then, the Bianchi constraint
transport equations (\ref{bianchi_constraint}) at $\tau=0$ are
satisfied up to order 6.
\end{lemma}
The proof of lemma \ref{fulfillment} proceeds, again, by direct
evaluation.

\section{Solving the transport equations}

The d-jets $J^{(6)}_{I^0}(v)$ and $J^{(6)}_{I^0}(\phi)$ described in
section 4.6 provide initial data for the transport equations of order
$p=1,\ldots,6$. Their solution would in turn yield the s-jets
$J^{(6)}_{I}(v)$ and $J^{(6)}_{I}(\phi)$ and thus the vector unknowns
$v$ and $\phi$ up to order $6$. Before describing these solutions, a
list of the assumptions being made is given.

\begin{assumptions}
In order to calculate the s-jets $J^{(6)}_{I}(v)$ and $J^{(6)}_{I}(\phi)$ it has been assumed that:
\begin{itemize}

\item[(i)] one has a time symmetric initial data set with an analytic
conformal metric $h_{\alpha\beta}$ ---respect to normal coordinates---
in a ball $B_a(i)$ of infinity;

\item[(ii)] the initial data for the conformal propagation equations
has been calculated in the cn-gauge in the way described in
section \ref{data_for_propagation};

\item[(iii)] the following choices for the function $\kappa$ appearing in the
  conformal factor $\Theta$ have been used: 
\[
\kappa=\rho, \qquad \kappa=\omega.
\]
\end{itemize}

\end{assumptions}  

\textbf{Remark.} It is noted that the assumption of analyticity of the
conformal compactification of the initial hypersurface is not strictly
needed. It is given here so for simplicity. It can be conveniently
substituted by the assumption that certain functions have Taylor
polynomials of a given order with decompositions in spherical
harmonics consistent with the expansion types.

The procedure to solve the transport equations (\ref{v_transport})
and (\ref{b_transport}) is extremely lenghty, but nevertheless
straightforward. Thus, it is quite amenable to a computer algebra treatment. A
``transport equations solver'' script in the computer algebra system {\tt
  Maple V} has been implemented for the calculations described in
\cite{Val04a}. The same scripts can be used, without the need of further
modifications, to perform the calculations required for the present
article. Because of the length of the expressions obtained, only qualitative
features of the solutions obtained are given. A detailed description of the
computer algebra implementation will be described elsewhere. 

Our first result is the following:
\begin{theorem}[solutions up to order 4]
Under the assumptions (i)-(iii), the solutions of the transport
equations for $p=0,\ldots,4$ are polynomial in $\tau$. Thus, they
extend smoothly to the sets $I^\pm$.
\end{theorem}

\textbf{Remark 1.} This last theorem is independent of the choice of the function $\kappa$. The polynomial expressions involved will nevertheless be different for each choice, but of the order of the polynomials is the same.
  
\textbf{Remark 2.} A complete listing of the solutions to the transport
equations up to order $3$ with the choice $\kappa=\rho$ is given in
\cite{FriKan00}. A description of the solutions of the Bianchi
transport equations for conformally flat data up to order $4$ with
$\kappa=\rho$ are given in \cite{Val04a}.

\bigskip
As it is to be expected from the results of \cite{Val04a}, the
situation is different when considering solutions to the transport
equations with $p\geq 5$.

\begin{theorem}[obstructions at order 5]
\label{thm:obst5}
Under the assumptions (i)-(iii), the solutions to the v-transport equations (\ref{v_transport}) at order $p=5$ are polynomial in $\tau$ and therefore smooth. On the other hand, the solutions to the $p=5$ Bianchi transport equations (\ref{b_transport}) are of the form
\[
\phi^{(5)}_j=\Upsilon^{(5)}\Big(f_j(\tau)\ln(1-\tau) +g_j(\tau)\ln(1+\tau)\Big) + h_j(\tau),
\]
for $j=0,\ldots,4$, where $f_j(\tau)$, $g_j(\tau)$ and $h_j(\tau)$ are
polynomials in $\tau$. The polynomials $f_j(\tau)$, $g_j(\tau)$ are
all of order $7$. In particular
\[
f_4(\tau)=g_0(-\tau)=(1-\tau)^3k(\tau)
\]
where $k(0)\neq 0$. Furthermore 
\[
\Upsilon^{(5)}=\sum_{k=0}^4 \Upsilon_k^{(5)}T_{4\phantom{k}2}^{\phantom{4}k},
\]
with
\begin{eqnarray}
&& \Upsilon_0^{(5)}=18mw_{2;4,0}-36\sqrt{6}w_{1;2,0}^2+r_{2;4,0}, \label{obs_50}\\
&& \Upsilon_1^{(5)}=18mw_{2;4,0}-72\sqrt{3}w_{1;2,0}w_{1,2,1}+r_{2;4,1}, \label{obs_51}\\
&& \Upsilon_2^{(5)}=18
mw_{2;4,2}-72w_{1;2,1}^2-72w_{1;2,0}w_{1;2,2}+r_{2;4,2}, \label{obs_52}\\
&& \Upsilon_3^{(5)}=18mw_{2;4,3}-72\sqrt{3}w_{1;2,1}w_{1;2,2}+r_{2;4,3}, \label{obs_53}\\
&& \Upsilon_4^{(5)}=18mw_{2;4,4}-36\sqrt{6}w_{1;2,2}^2+r_{2;4,4}. \label{obs_54} 
\end{eqnarray}

\end{theorem}

Thus, as a consequence of this theorem, the solutions of the $p=5$
Bianchi transport equations are ---in general--- not smooth
(i.e. $C^\infty$), but merely of class $C^2$ at the sets
$I^\pm$. Because of this, the quantities $\Upsilon_j^{(5)}$,
$j=0,\ldots,4$ will be referred as to the \emph{order $5$
obstructions}. There are several remarks that come now into place.

\textbf{Remark 1.} Firstly, note that if one sets $r_{2,4,k}=0$,
$k=0,\ldots 4$ one recovers the order $5$ obstructions found in
\cite{Val04a} for conformally flat data.

\textbf{Remark 2.} For conformally flat initial data the obstructions
coincide ---modulo some irrelevant numerical factor--- with the
Newman-Penrose constants of the time development. For the developments
of non-conformally flat time symmetric data  the
Newman-Penrose constants can be expressed in terms of the initial data
by
\begin{eqnarray*}
&& G_0^{(5)}=18mw_{2;4,0}-36\sqrt{6}w_{1;2,0}^2-\frac{1}{1016}r_{2;4,0},\\
&& G_1^{(5)}=18mw_{2;4,0}-72\sqrt{3}w_{1;2,0}w_{1,2,1}-\frac{1}{1016}r_{2;4,1}, \\
&& G_2^{(5)}=18
mw_{2;4,2}-72w_{1;2,1}^2-72w_{1;2,0}w_{1;2,2}-\frac{1}{1016}r_{2;4,2}, \\
&& G_3^{(5)}=18mw_{2;4,3}-72\sqrt{3}w_{1;2,1}w_{1;2,2}-\frac{1}{1016}r_{2;4,3}, \\
&& G_4^{(5)}=18mw_{2;4,4}-36\sqrt{6}w_{1;2,2}^2-\frac{1}{1016}r_{2;4,4}.  
\end{eqnarray*}
 
Thus, we see that in general for time symmetric constants the order
$5$ obstructions are different to the Newman-Penrose constants of the
time development. The coincidence occurring in the case of conformally
flat data lead to conjecture that because static solutions should have
no logarithmic terms in their expansions then their Newman-Penrose
constants should be zero. Our results show that this is also in
general not the case. Explicit calculations using the Weyl class of
solutions by S. Dain have lead to the same conclusion \cite{Dai02}.

\textbf{Remark 3.} From the way lower order solutions to the transport
equations feed into higher order transport equations that if the
obstructions (\ref{obs_50})-(\ref{obs_54}) do not vanish then the
solutions $v^{(6)}$ will already contain terms of the form
$\ln(1\pm\tau)$, and even more, $\phi^{(6)}$ will contain
$\ln^2(1\pm\tau)$ terms. The situation is bound to be even worse for
$p\geq 7$. The hyperbolic nature of the propagation equations implies
in turn that the non-smoothness of the solutions of the transport
equations at $I^\pm$ will be propagated along the generators of null
infinity. However, a proper discussion of this phenomenon is still to
be made.

\bigskip
Our investigation is focused on solutions to the transport equations
which are consistent with a time development of the initial with
smooth null infinity. Therefore, the situation of most interest for us
is when the so-called order $5$ obstructions vanish. The consideration
of order $6$ transport equations under this assumption leads to the
following result.

\begin{theorem}[obstructions at order 6]
\label{thm:obst6}
Under the assumptions (i)-(iii) and if the order $5$ obstructions (\ref{obs_50})-(\ref{obs_54}) vanish, the solutions to the v-transport equations (\ref{v_transport}) at order $p=6$ are polynomial in $\tau$. The solutions to the $p=6$ Bianchi transport equations (\ref{b_transport}) are of the form
\[
\phi^{(6)}_j=\Upsilon^{(6)}\Big(\hat{f}_j(\tau)\ln(1-\tau) +\hat{g}_j(\tau)\ln(1+\tau)\Big) + \hat{h}_j(\tau),
\]
for $j=0,\ldots,4$, where $\hat{f}_j(\tau)$, $\hat{g}_j(\tau)$ and
$\hat{h}_j(\tau)$ are again polynomials in $\tau$. The polynomials
$\hat{f}_j(\tau)$, $\hat{g}_j(\tau)$ are of order $9$, and in
particular
\[
\hat{f}_4(\tau)=\hat{g}_0(-\tau)=(1-\tau)^4\hat{k}(\tau),
\]
where $\hat{k}(0)\neq0$. Furthermore
\[
\Upsilon^{(6)}=\sum_{k=0}^6 \Upsilon_k^{(6)}T_{6\phantom{k}3}^{\phantom{6}k},
\]
with
\begin{eqnarray}
&&\!\!\!\!\!\!\!\!\!\!\!\!\!\!\!\!\!\! \Upsilon_0^{(6)}=24m^2w_{3;6,0}-48\sqrt{15}mw_{1;2,0}w_{2;4,0}+r_{3;6,0}, \label{obs_60}\\
&&\!\!\!\!\!\!\!\!\!\!\!\!\!\!\!\!\!\! \Upsilon_1^{(6)}=24m^2w_{3;6,1}-48\sqrt{10}mw_{1;2,0}w_{2;4,1}-48\sqrt{5}mw_{1;2,1}w_{2;4,0}+r_{3;6,1},\\
&&\!\!\!\!\!\!\!\!\!\!\!\!\!\!\!\!\!\! \Upsilon_2^{(6)}=24m^2w_{3;6,2}-72\sqrt{6}m w_{1;2,0}w_{2;4,2} -72\sqrt{2}mw_{1;2,1}w_{2;4,1}+r_{3;6,2},\\
&&\!\!\!\!\!\!\!\!\!\!\!\!\!\!\!\!\!\! \Upsilon_3^{(6)}=24m^2w_{3;6,3}-144m w_{2;4,2}w_{1;2,1}-48\sqrt{3}mw_{1;2,0}w_{2;4,3}-48\sqrt{3}m w_{1;2,2}w_{2;4,1}+r_{3;6,3},\\
&&\!\!\!\!\!\!\!\!\!\!\!\!\!\!\!\!\!\! \Upsilon_4^{(6)}=24m^2w_{3;6,4}-72m\sqrt{6} w_{1;2,2} w_{2;4,2}-72\sqrt{2}m w_{1;2,1}w_{2;4,3}+r_{3;6,4},\\
&&\!\!\!\!\!\!\!\!\!\!\!\!\!\!\!\!\!\! \Upsilon_5^{(6)}=24m^2w_{3;6,5}-48\sqrt{10}mw_{1;2,2}w_{2;4,3}-48\sqrt{5}mw_{1;2,1}w_{2;4,4}+r_{3;6,5},\\
&&\!\!\!\!\!\!\!\!\!\!\!\!\!\!\!\!\!\! \Upsilon_6^{(6)}=24m^2w_{3;6,6}-48\sqrt{15}mw_{1;2,2}w_{2;4,4}+r_{3;6,6}. \label{obs_66}
\end{eqnarray}
\end{theorem}

Again, if one sets the terms associated with the Ricci scalar,
$r_{2;4,k}$, $r_{3;6,k}$ to zero, one recovers the obstructions obtained in
\cite{Val04a} for conformally flat initial data.

It is mentioned by passing that it is quite likely that the order $5$
and order $6$ obstructions (\ref{obs_50})-(\ref{obs_54}) and
(\ref{obs_60})-(\ref{obs_66}) are associated to some kind of conserved
quantities at null infinity. Using an analysis along the lines of the
asymptotic characteristic initial value problem, it was shown in
\cite{ChrMacSin95} that there are certain conserved quantities
---besides the so-called logarithmic Newman-Penrose constants---
associated with the first logarithmic terms appearing in their
expansions. In the basis of our results we speculate that these
conserved constants are related with our obstructions, although a
discussion of this will also be left for the future.

The complexity of the expressions involved preclude us from a direct
evaluation of the order $7$ and higher solutions to the transport
equations. In order to reduce this complexity one can consider data with axial
symmetry. In this way it is possible to complete the expansions up to order
$8$ inclusive. Due to the axial symmetry, there is only one obstruction at
every order. These are:
\begin{eqnarray*}
&& \Upsilon^{(7)}_4=5040mw_{4;8,4}-23040w_{1;2,1}w_{3;6,3}-25920w_{2;4,2}^2-161r_{4;8,4}, \\
&& \Upsilon^{(8)}_5=1440mw_{5;10,5}-4800w_{1;2,1}w_{4;8,4}-19200w_{2;4,2}w_{3;6,3}+37r_{5;10,5}.
\end{eqnarray*}

\section{The obstructions and the static initial data}

In order to provide an interpretation of the obstructions obtained in
the previous section, and in view of the results of \cite{Val04a}, let
us consider for a moment static initial data given in the
cn-gauge. The static data gives rise to time developments with a
smooth null infinity \cite{Dai01b}. Thus, one should expect the
obstructions should automatically vanish for this kind of data.

The line element of static solutions to the field equations can be written, in
the physical spacetime in the form
\[
\widetilde{g}=\psi^2dt^2+\widetilde{h}
\]
where $\widetilde{h}$ is a $t$-independent, negative definite,
3-metric, and $\psi$ is its $t$-independent norm,
$\psi=\sqrt{\widetilde{h}(k,k)}$ of the Killing vector field
$k=\partial_t$. The static vacuum field equations are then given by
\begin{eqnarray*}
&& \widetilde{r}_{\alpha\beta}=\frac{1}{\psi} \widetilde{D}_\alpha \widetilde{D}_\beta \psi, \\
&& \Delta_{\widetilde{h}}\psi=0,
\end{eqnarray*}
 where $\widetilde{r}_{\alpha\beta}$, $\widetilde{D}$,
 $\Delta_{\widetilde{h}}$ are respectively the Ricci tensor, the
 covariant derivative and the Laplacian with respect to the metric
 $\widetilde{h}$. Now, introducing the rescaling
 $h_{\alpha\beta}=\Omega^2\widetilde{h}_{\alpha\beta}$ where $\Omega$
 is the conformal factor arising from the Licnerowicz equation
 (\ref{licnerowicz_equation}), contracting with the frame $e_{(i)}$,
 $i=1,2,3$ and then using the spatial Infeld symbols $\sigma^{(i)}_{ab}$
 one arrives to the following (space spinorial) form of the static equations
\begin{eqnarray*}
&& s_{abcd}+\frac{1}{3}r h_{abcd}+\frac{1}{\Omega}D_{ab}D_{cd}\Omega +h_{abcd}\left(\frac{1}{\Omega}\Delta_h\Omega -\frac{2}{\Omega^2}D^{ef}D_{ef}\Omega\right) \label{static_1}\\
&&  \phantom{XXXXXXXXXX}= \frac{1}{\psi}\left( D_{ab}D_{cd}\psi + \frac{2}{\Omega} D_{(ab}D_{cd)}\Omega -\frac{1}{\Omega}h_{abcd} D^{ef}\psi D_{ef}\Omega\right), \\
&& \Delta_h \left(\Omega^{-1/2}\psi \right)=\frac{1}{8}r\Omega^{-1/2}\psi. \label{static_2}
\end{eqnarray*}
Given a time
symmetric initial data set in a the cn-gauge ---and following previous
discussions--- we will say that the initial data set is \emph{static
up to order $p$} if given the corresponding order $p$ d-jets
$J^{(p)}_{I^0}(v)$, $J^{(p)}_{I^0}(\phi)$, there is a static solution
with order $p$ d-jets $J^{(p)}_{I^0}(v_s)$, $J^{(p)}_{I^0}(\phi_s)$
such that
\[
J^{(p)}_{I^0}(v)=J^{(p)}_{I^0}(v_s), \qquad J^{(p)}_{I^0}(\phi)=J^{(p)}_{I^0}(\phi_s).
\]

\bigskip
Substituting the expansions for $s_{abcd}$, $\Omega$, $r$ and the
connection and frame associated with $D_{ab}$ found in section \S4, one
obtains after some {\tt Maple V} calculations the following lemma.
\begin{lemma}
Given the assumptions (i)-(iii), necessary and sufficient conditions for
a time symmetric initial data set to be static up to order 3 are that
the obstructions (\ref{obs_50})-(\ref{obs_54}) and
(\ref{obs_60})-(\ref{obs_66}) vanish.
\end{lemma}

Combining theorems \ref{thm:obst5} and \ref{thm:obst6} together with
the previous lemma, we have the proved our main result, and thus
provided an interpretation of the obstructions
(\ref{obs_50})-(\ref{obs_54}) and (\ref{obs_60})-(\ref{obs_66}).

\begin{main}[Main theorem, technical version]
Necessary condition for the s-jets $J_I^{(7)}(v)$ and
$J_I^{(6)}(\phi)$ to extend smoothly to the sets $I^\pm$ is that the
initial data set is static up to order 3.
\end{main}

Higher order results along the same
lines can be obtained if one restricts the analysis to initial data
which is axially symmetric. All together, these results provide what
we believe is fair evidence for the general
conjecture for time symmetric data put forward in the introduction.

\section{Expansions for the Bondi mass near spatial infinity}

It is of interest to see how the results of the previous section can be obtain
by means of a different kind of arguments. Namely, we will show how one can
relate the obstructions obtained in the previous sections to
the Bondi mass. It is intuitively clear ---although its verification requires
some calculations, see e.g. \cite{Fri03c}--- that if one starts with a Cauchy
problem near spatial infinity as the one described here, in which the initial
data is static, the Bondi mass should be constant and equal to the ADM mass of
the spacetime. The converse, namely starting from a regular finite Cauchy
problem at spatial infinity with the additional requirement that the Bondi
mass should be constant to deduce the existence of a static Killing vector
near spatial infinity, is also intuitively plausible but much more
complicated, requiring an existence theorem for the Cauchy problem in the
neighbourhood of the sets $I^\pm$ which is not yet available.

The standard way of introducing the Bondi mass and other quantities
defined at null infinity ---like the Newman-Penrose constants--- is to
construct an \emph{ad hoc} coordinate system and frame well adapted to
the geometry of spacetime near the intersection of null infinity with
an outgoing light cone.  We will refer to these coordinates and the
concomitant adapted frame to as \emph{the Newman-Penrose (NP)
gauge}. A detailed discussion on how to construct this gauge can be
found in \cite{FriKan00}. In the NP gauge ---and using the NP
formalism notation-- the Bondi mass is given by
\begin{equation}
\label{bondi_mass}
m_B=-\frac{1}{4\pi}\oint \left(\psi_2^0-\sigma_0\dot{\overline{\sigma}}_0\right) dS,
\end{equation}
where the functions $\psi_2^0$ and $\sigma_0$ ---depending on a
retarded time $u$ and the angular dependence-- are the leading terms
of the component $\widetilde{\psi}_2$ of the Weyl tensor and the
spin coefficient $\sigma$ ---the shear. These quantities are given in
the physical space time, and for the class of solutions of the Einstein
equations under consideration it can be seen that
\[
\widetilde{\psi_2}=\psi_2^0 \Theta^3+\cdots, \qquad \widetilde{\sigma}=\sigma_0\Theta^2+\cdots.
\]
The integral in (\ref{bondi_mass}) is evaluated over the cut of
$\scri^+$ defined by $u=const.$ The relevance of the Bondi mass is
that under suitable assumptions it can be shown to be non-negative and
non-increasing, i.e. $\dot{m}_B=\partial_u m_B=0$. Furthermore, under
some extra assumptions it can be shown to tend to the ADM mass of the
spacetime as one approaches spatial infinity ---see
\cite{AshMag79,FriKan00,Val03b}. It is worth pointing that all these
discussions about the limit of the Bondi mass contain \emph{fide}
assumptions that shall eventually be removed once a complete theory of
spatial infinity is achieved.

A major inconvenient of the Bondi mass in the form given by formula
(\ref{bondi_mass}) is the gauge in which it is given, which although
very convenient for theoretical discussions, hampers its evaluation in
direct examples. The connection between the NP gauge and the gauge
used in the present article ---sometimes refered as to the F gauge---
has been obtained in \cite{FriKan00}. The transformation between
gauges involves the determination of a conformal factor $\theta$, such
that $\theta\Theta$ is the conformal factor in the NP gauge, and a
rotation of the frame $c_{aa'}$ introduced in section 2, and which is
realized by matrices $\Lambda^{a}_{\phantom{a}b}\in
SL(2,\Complex)$. Under the choice, $\kappa=\omega$, of the function
$\kappa$ appearing in the conformal factor given in
equation(\ref{st_cf}), the locus of null infinity is given in a simple
way by $\tau=\pm 1$. Hence, the coordinate $\rho$ can be used to
parametrise the Bondi mass.

Instead of calculating the Bondi mass using the formula
(\ref{bondi_mass}), we will make use of the fact that
\begin{equation}
\label{m_dot}
\dot{m}_B= -\frac{1}{4\pi} \oint \dot{\sigma}_0\dot{\overline{\sigma}}_0 dS,
\end{equation}
and that
\[
\psi_3^0=iX_+\dot{\overline{\sigma}}_0, 
\]
where $\widetilde{\psi}_3=\psi_3^0\Theta^2+\cdots$ ---see
e.g. \cite{Ste91}. The coefficient $\psi_3^0$ has spin weight 1, and
thus it follows that in this case the operator $X_+$ has a uniquely defined
inverse $X_+^{-1}$ ---see e.g. \cite{ExtNewPen69}. Now, from the
discussion given in \cite{FriKan00} to calculate the Newman-Penrose
constants in the F gauge, it also follows that
\[
\widetilde{\psi}^0_3=\frac{1}{\theta}\Lambda^a_{\phantom{a}0}\Lambda^b_{\phantom{b}1}\Lambda^c_{\phantom{c}1}\Lambda^d_{\phantom{d}1}\phi_{abcd}.
\]
Now, only terms
with $T_{0\phantom{0}0}^{\phantom{0}0}$ survive this integration of equation
(\ref{m_dot}). A final integration along the generators of $\scri^+$ yields
the following result.

\begin{theorem}[Expansions for the Bondi mass]
Under the assumptions (i)-(iii) one has that
\[
m_B=m-K\sum_{k=0}^4 |\Upsilon^{(5)}_k|^2 \rho^7+\O(\rho^8),
\] 
where $K$ is a numeric factor. If $\Upsilon^{(5)}_k=0$, $k=0,\ldots,4$ then
\[
m_B=m-K'\sum_{k=0}^6 |\Upsilon^{(6)}_k|^2 \rho^9+\O(\rho^{10}),
\]
where $K'$ is a numeric factor. If $\Upsilon^{(6)}_k=0$, $k=0,\ldots,6$ then
\[
m_B=m+\O(\rho^{11}).
\]
\end{theorem}
These results are the natural extension of those presented in \cite{Val03b}.

\section{Conclusions}
The results presented in this article constitute what we believe is
fair evidence to conjecture that the only time reflexion symmetric
spacetimes ---i.e. those arising as the developments of time symmetric
initial data sets--- with a smooth conformal extension to null
infinity are those that are static in a neighbourhood of spatial
infinity. 

We would like to contrast the situation of the gravitational field with that
of the spin 2 zero rest mass field on a Minkowski background. As shown in
\cite{Fri03b,Val03a}, an analogue of the regularity condition
(\ref{reg_cond}) is a necessary and sufficient condition to obtain
logarithm-free solutions to the transport equations, therefore ensuring that
the field is smooth at null infinity ---the analogue of Penrose's proposal for
linear fields. It would be interesting to analyse what happens in the case of
fields propagating on curved backgrounds.

The physical intuition suggests that staticity near spatial infinity
is also a statement about the behaviour of the gravitational field
---and their sources--- at very early and late times ---that is, near
$i^\pm$. This is easier to see in the case of initial data which are
Schwarzschildean near spatial infinity. The vanishing of the
Newman-Penrose constant implies then in turn the vanishing of the Weyl
tensor at $i^\pm$ as discussed in the introduction. The interpretation
of data which are static near spatial infinity is in any case not that
clear for the non-conformally flat case. For as it has been seen,
their Newman-Penrose constants do not vanish. Nevertheless, the fact
that it is possible to make ---non-trivial--- statements about the
behaviour of spacetimes at $i^\pm$ without having to solve the
equations is already astonishing. It is worth mentioning that recent
results by Chru\'sciel \& Delay \cite{ChrDel03} make possible now to
construct initial data sets of the form required by the conjecture
presented in this article ---i.e. static near $i^0$.

Finally, answering the question raised in the title of this article,
asymptotic simplicity does seem to exclude gravitational radiation near
spatial infinity. A rigorous proof of the conjecture
raised here awaits further, deeper, insights into the structure of the
(conformal) Einstein field equations.

\subsection*{Acknowledgements} 
I would like to thank H. Friedrich who introduced me to this research
topic and has provided me with invaluable advice. I also acknowledge
several enrichening and helpful discussion with R. Beig, S. Dain and
J. Winicour. I also thank an anonymous referee for a careful reading
of the manuscript and an important observation leading to lemma 4.

This work is funded by a Lise Meitner fellowship (M690-N02 and
M814-N02) of the Fonds zur Forderung der Wissenschaftlichen Forschung
(FWF), Austria. The computer algebra calculations here described have
been performed in the computer facilities of the Albert Einstein
Institute, Max Planck Institute f\"ur Gravitationsphysik, Golm,
Germany.


\end{document}